\documentclass[aps,pre,reprint,superscriptaddress,amsmath,amssymb,longbibliography,onecolumn]{revtex4-2}
\usepackage{graphicx}
\usepackage{physics}
\usepackage{xr}
\usepackage{verbatim}
\usepackage{color,xcolor}
\usepackage{mathtools}
\usepackage{orcidlink}
\usepackage{amsmath}
\usepackage{amssymb}
\definecolor{Red}{RGB}{200,35,10}
\definecolor{Blue}{RGB}{15,75,210}
\definecolor{Green}{RGB}{5,140,80}

\newcommand{\fght}{fGHT }
\newcommand{\mght}{mGHT }

\begin{document}
\title{Beyond average: heterogeneous first-passage dynamics in many-particle systems with resetting}
%
\author{Juhee Lee\orcidlink{0000-0003-3318-6377}}
\affiliation{Integrated Science Lab, Department of Physics, Ume{\aa} University, Ume{\aa}, Sweden}
\author{Seong-Gyu Yang\orcidlink{0000-0003-0583-0006}}
\affiliation{Integrated Science Lab, Department of Physics, Ume{\aa} University, Ume{\aa}, Sweden}
\affiliation{Department of Physics and Institute of Basic Science, Sungkyunkwan University, Suwon 16419, Republic of Korea}
\author{Ludvig Lizana\orcidlink{0000-0003-3174-8145}}
\affiliation{Integrated Science Lab, Department of Physics, Ume{\aa} University, Ume{\aa}, Sweden}
\date{\today}
\begin{abstract}
Stochastic resetting is well understood for single-particle first-passage processes, but its consequences for collective first-passage behavior remain less clear. We address this problem in a many-particle system where all particles reset to the position of the rightmost particle, a protocol motivated by problems in artificial selection and avoidance. We use stochastic simulations of particles diffusing in a confining potential with an adsorbing boundary to examine two notions of group arrival: the first group-hitting time, when the first particle reaches the boundary, and the median group-hitting time, when half of them have reached it. We find that resetting produces broad hitting-time distributions with extended plateaus spanning several orders of magnitude. As the resetting rate increases, these plateaus extend and the mean group-hitting times grow rapidly. At the same time, the first-passage dynamics become increasingly heterogeneous. Consequently, the most probable and mean hitting times become widely separated, indicating the absence of a single characteristic time scale. These results demonstrate that the definition of group arrival is crucial for understanding and controlling collective first-passage behavior under resetting.
\end{abstract}
\maketitle
%
%
\section{Introduction}
Stochastic resetting has attracted considerable attention as a paradigm for nonequilibrium search processes~\cite{evans2011diffusion,pal2017first,chechkin2018random,durang2019first} and avoidance problems~\cite{de2020optimization,de2021optimization}.
Most of these studies focus on single-particle dynamics.
More recently, several studies have explored resetting in many-particle systems and in group search~\cite{evans2020stochastic,nagar2023stochastic,evans2022exactly,mercado2018lotka,da2018interplay,miron2021diffusion,biroli2023extreme,sasorov2023probabilities,grange2020non,biroli2023critical,singh2022capture,singh2023bernoulli}.
For example, Ref.~\cite{biroli2023extreme} studied the nonequilibrium steady state of a group of Brownian particles that reset to their common initial position and showed that several observables can be computed exactly.
Resetting protocols in many-particle systems are typically divided into two classes: global and local.
In global resetting, all particles reset simultaneously~\cite{evans2020stochastic,nagar2023stochastic,evans2022exactly,mercado2018lotka,miron2021diffusion,biroli2023extreme,biroli2023critical,singh2022capture,singh2023bernoulli}, whereas in local resetting, each particle resets independently~\cite{nagar2023stochastic,da2018interplay,grange2020non,vilk2022fluctuations,krapivsky2022competition,siboni2021fluctuations,biroli2023critical,sasorov2023probabilities}.
Another important distinction is how the resetting position is determined.
Most studies consider resetting to a predefined state, such as the origin~\cite{evans2020stochastic,nagar2023stochastic,evans2022exactly,mercado2018lotka,da2018interplay,miron2021diffusion,biroli2023extreme,sasorov2023probabilities} or the initial position~\cite{evans2020stochastic,nagar2023stochastic,grange2020non,biroli2023critical,singh2022capture,singh2023bernoulli}.
More dynamic protocols include resetting by a random amplitude~\cite{dahlenburg2021stochastic}, partial resetting~\cite{tal2022diffusion}, and resetting to the maximum position reached in the past~\cite{majumdar2015random}.

To complement these studies, we recently developed a resetting protocol based on the extreme-value statistics of the group~\cite{lee2026general}.
At a fixed rate, all particles are reset to the position of the rightmost particle relative to an undesired region, although the theory applies more generally.
This choice is motivated by problems in bacterial evolution under selection pressure~\cite{davies2010origins,neu1992crisis,munita2016mechanisms,blair2015molecular,lee2025success}.
One prominent example is antibiotic resistance, which represents a genetically stable but undesired state.
By growing several bacterial subpopulations and applying artificial selection, one can steer the population away from such a state by repeatedly selecting the least-fit subpopulation.
This procedure effectively resets the bacterial population in trait space, with the resetting position determined by the extreme-value statistics of the population's ecological dynamics.

While our previous study derived several numerical and analytical results for extreme-value group resetting~\cite{lee2026general}, it only quantified avoidance probabilities using stationary observables, such as the mean position of the group's center of mass and the squared coefficient of variation.
Here, we investigate first-passage dynamics and determine how the group-hitting times depend on group size, resetting rate, initial position, and potential strength.
We also develop a simple analytical theory that captures the main features of the group-hitting-time distributions.
We find broad distributions spanning several orders of magnitude, with the mean becoming increasingly unrepresentative of individual realizations as the resetting rate increases.
These results highlight the need to look beyond average first-passage times to characterize collective first-passage dynamics~\cite{mattos2012first,godec2016first}.

\section{Model}
%
\begin{figure}[t!]
    \centering\includegraphics{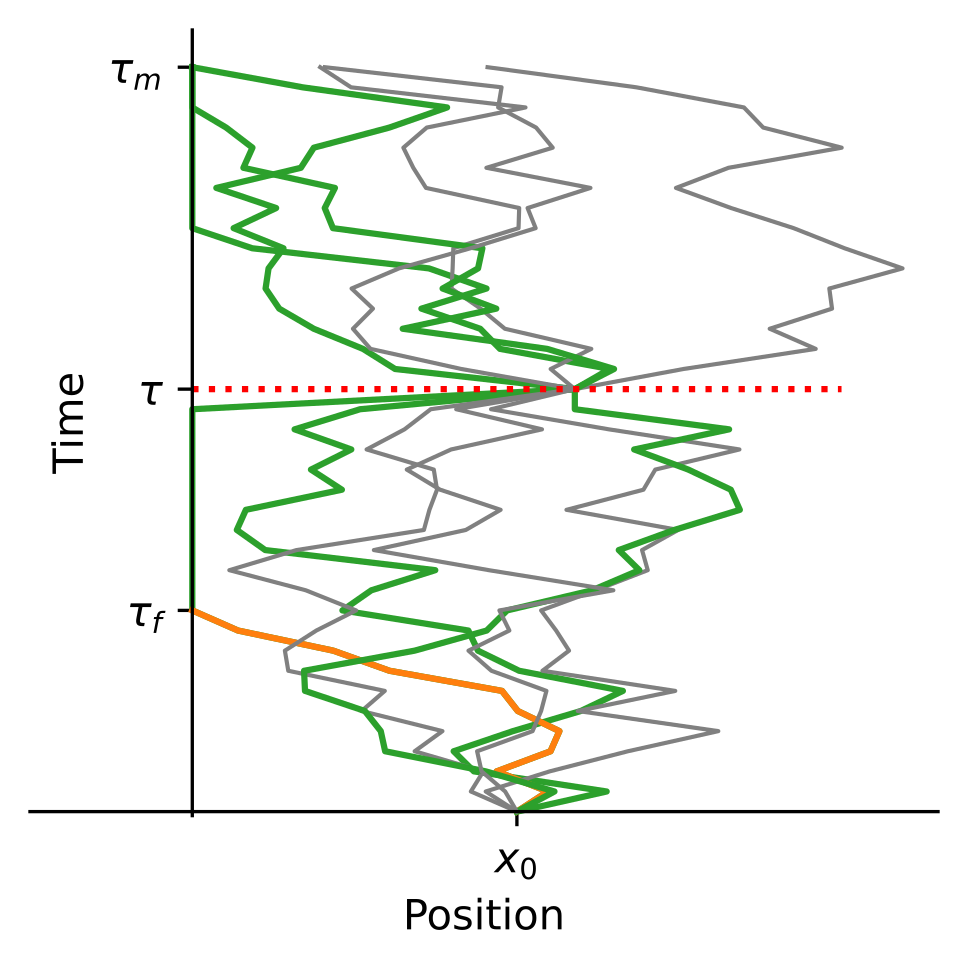}%

    \caption{\label{fig:1}
    Schematic of the resetting dynamics with adsorbing boundary (left vertical line) for $N=6$.
    The particles diffuse from $x_0$ and relocate to the rightmost particle's position at $\tau$.
    The first particle hits the boundary at $\tau_f$, and half of them at $\tau_m$.
    }
\end{figure}

We consider $N$ overdamped Brownian particles with coordinates $x_i\ (i=1,2,\cdots,N)$ moving in a one-dimensional harmonic potential $V(x)=\frac{1}2kx^2$ with an adsorbing boundary at $x=0$ between consecutive resetting events.
All particles start from the initial position $x_i(t=0)=x_0 (>0)$ and diffuse with diffusion coefficient $D$ until they reach the boundary.
The diffusion dynamics of each particle $i$ follows the Langevin equation 
\begin{equation}
dx_i(t) = -V'(x_i) dt+\sqrt{2D}\ dW_i(t),
\label{eq:1}
\end{equation}
where $W_i(t)$ denotes a Wiener process, and the prime in the potential denotes the first derivative.
If a particle crosses the boundary (i.e., $x_i \leq 0$), its position remains fixed at $x_i=0$ while the others are diffusing.
In addition to Brownian motion, all particles, including those at $x=0$, undergo a reset with probability $r dt$, where $r$ is the resetting rate and $dt$ is a short time interval.
When the reset occurs, all particles simultaneously and instantaneously relocate to the rightmost particle's position, $\max\lbrace x_i\rbrace$, and the Brownian motion continues from there according to Eq.~\eqref{eq:1}.

Furthermore, unlike single-particle resetting, many-particle resetting admits several rules for defining the group's first arrival.
We focus on two adsorption criteria.
The first one is satisfied as soon as any single particle reaches $x=0$ for the first time.
We denote this First Group-Hitting Time (fGHT), or $\tau_f$.
The second criterion is a majority rule.
It triggers when at least half the particles are adsorbed at $x=0$.
We call this Median Group-Hitting Time (mGHT), or $\tau_m$.

To illustrate our system's dynamics and the adsorption criteria, we show $N=6$ typical particle trajectories in Fig.~\ref{fig:1}.
Starting at position $x=x_0$, all particles diffuse and drift to the left because of the harmonic potential.
At time $t=\tau$,  all suddenly reset their positions to that of the rightmost particle, and the particles continue to diffuse from there.
At time $t=\tau_f$, one of the particles (orange) reaches $x=0$, and satisfies the \fght criterion.
But for the other criterion, at least half of the group ($3$ particles) must have reached the boundary.
In that case, the first particle remains at the boundary (green vertical line after orange line) until resetting event or two more arrive (green lines).
This happens at $t=\tau_m$, at which point \mght is satisfied.

\subsection{Simulations}
We use the Euler-Maruyama method~\cite{kloeden1992stochastic,gardiner2009stochastic} to obtain particle trajectories from Eq.~\eqref{eq:1}.
Throughout this study, we fix the parameters as $x_0=1$, $k=1$, and $D=2$ to investigate the effect of group resetting unless stated otherwise.
Furthermore, we adjust the time step based on $r$: $dt=0.01$ for $r\leq 10$, and $dt=0.1/r$ for $r>10$.
This achieves at least a 10\% chance of resetting per time step.
We set the maximum time to $t_\text{max}=10^5$, and the number of simulation runs to $10^5$ ("sample size").

\section{Results}

\begin{figure*}
 \centering
 \includegraphics[width=\columnwidth]{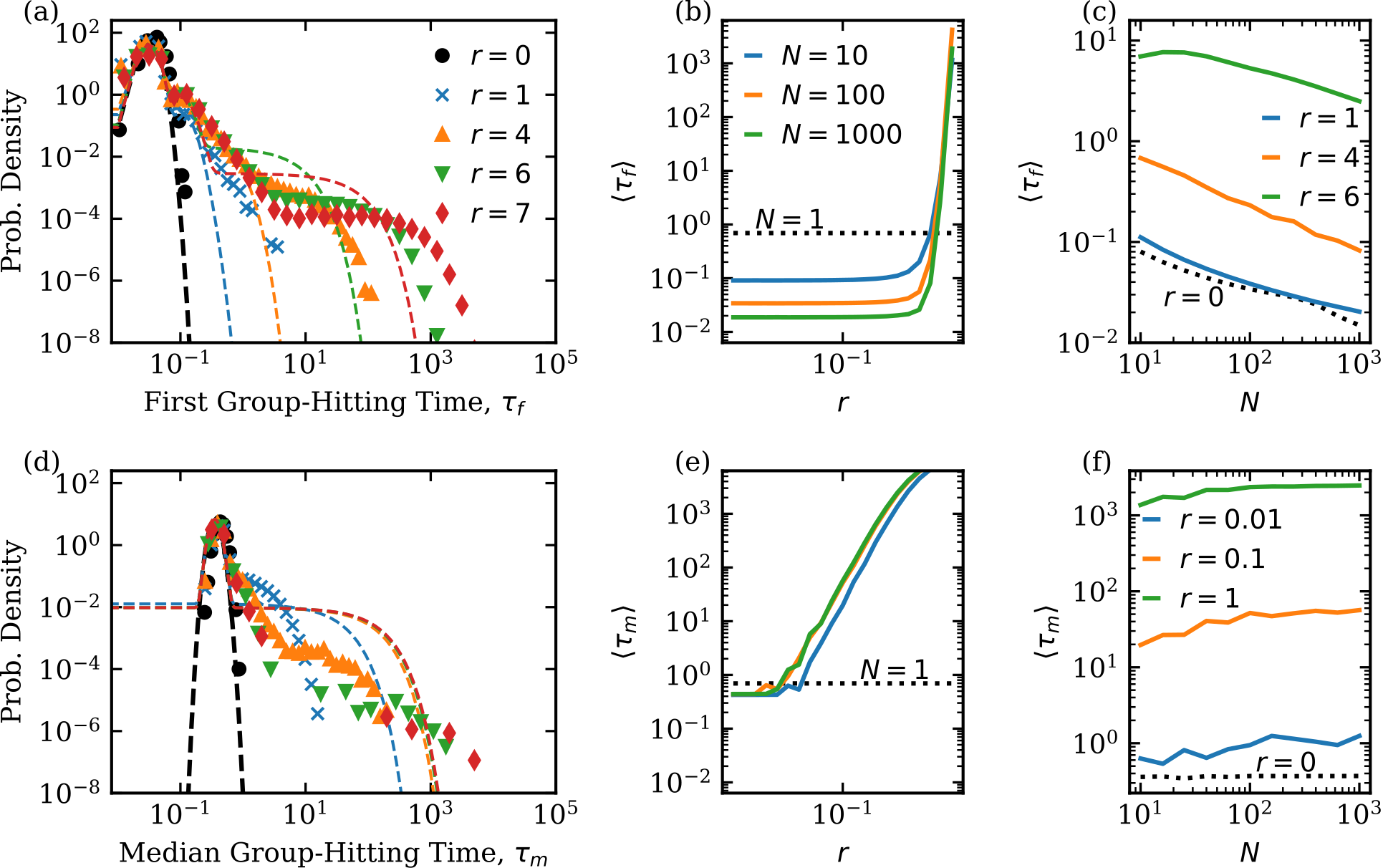}%
 \caption{\label{fig:2} 
    First ($\tau_f$) and median ($\tau_m)$ group-hitting time distributions for $N=100$ [(a) and (d)].
    Symbols denote simulations and dashed lines indicate theoretical approximations, Eq.~\eqref{eq:appr}.
    Each color corresponds to a different resetting rate $r$.
    We note that the distribution tails extend as $r$ increases.
    Both (ensemble) averages $\langle \tau_f \rangle$ and $\langle \tau_m \rangle$ grow sharply for large $r$ [(b) and (e)], but behave in the opposite way with respect to $N$: $\langle \tau_f \rangle$ decreases at sufficiently large $N$ while $\langle \tau_m \rangle$ increases for all $N$[(c) and (f)].
    Parameters: $D=2$,\ $k=1$.}
\end{figure*}

\subsection{Hitting time distributions are broad}

We display the \fght and \mght distributions in Figs.~\ref{fig:2}(a) and (d), respectively, alongside the case without reset $r=0$ (dashed black lines).
We note that both distributions follow the no-reset behavior for short group hitting times. 
At later times ($\tau_f\approx 10^{-1}$ and $\tau_m\approx 10^0$), the curves start to deviate from the dashed line and become heavier.
For sufficiently large $r$, the tail turns into a plateau that extends over several orders of magnitude.
For $\tau_m$ and $r\gtrsim6$, the plateau is separated from the early-time peak by a sparsely populated region around $\tau_m\sim10^1$.
To test whether this apparent gap results from limited sampling, we performed additional simulations with a hundredfold larger sample.
The region remains sparsely populated in these simulations, suggesting that the gap reflects an intrinsically low probability of intermediate group-hitting times rather than insufficient sampling (see Appendix~\ref{appendix:d}).

As the plateaus extend with $r$, the mean group-hitting times increase.
We show $\langle\tau_f\rangle$ and $\langle\tau_m\rangle$ as functions of $r$ for three group sizes ($N=10,10^2,10^3$) in Figs.~\ref{fig:2}(b) and (e).
Because the simulations terminate at $t_{\max}=10^5$, some realizations do not satisfy the corresponding hitting criterion within the simulation time.
We therefore obtain a lower-bound estimate of the means by assigning the right-censored realizations the value $t_{\max}$:
\begin{equation}
\langle\tau_{f,m}\rangle
=(1-S(t_{\max}))\langle\tau_{f,m}\rangle_{\mathrm{obs}}
+S(t_{\max})t_{\max},
\end{equation}
where $\langle\tau_{f,m}\rangle_{\mathrm{obs}}$ is the mean over realizations that terminate before $t_{\max}$, and $S(t_{\max})$ is the fraction that survive until $t_{\max}$.
Since the actual hitting times of these surviving realizations exceed $t_{\max}$, this procedure provides a lower bound on the mean.
Accordingly, as $S(t_{\max})\to 1$ as $r\to \infty$, the estimate approaches $\langle\tau_{f,m}\rangle \to t_{\max}$, even though the true mean can be larger.

For small $r$, both mean group-hitting times depend only weakly on the resetting rate, but their dependence on group size is different.
The mean \fght decreases with increasing $N$, whereas the mean \mght is comparatively insensitive to $N$.
As $r$ increases, both mean hitting times rise rapidly.
In this regime, the \fght curves for different $N$ converge, while the \mght curves become more separated.
The $N=1$ results (black dashed lines) are included as a reference case without group resetting.
At large $r$, the lower-bound estimates approach $t_{\max}$ as an increasing fraction of realizations remains right-censored.

To better understand the group-size dependence, we plot $\langle \tau_f \rangle$ and $\langle \tau_m \rangle$ against $N$ for three different resetting rates, $r=1,\ 4,\ 6$ and $r=0.01,\ 0.1,\ 1$, respectively [Figs.~\ref{fig:2}(c) and (f)], including $r=0$ as a reference.
The two group-hitting times show opposite dependences on $N$: $\langle\tau_m\rangle$ increases with $N$, whereas $\langle\tau_f\rangle$ decreases.
For \mght, increasing $N$ makes unusual trajectories more likely, because it increases the chance that one of them makes an unusually large excursion to the right of the previous resetting position.
Since the group resets to the rightmost particle, such extreme excursions shift subsequent resetting positions farther from the adsorbing boundary.
This reduces the probability that half of the particles reach $x=0$ and thereby increases $\langle\tau_m\rangle$.
The $N$-independent behavior for $r=0$ in Fig.~\ref{fig:2}(f) supports the interpretation that this increase originates from the group-resetting mechanism.

Increasing $N$ introduces a competing effect for \fght.
Since \fght is determined by the first particle to reach the adsorbing boundary, a larger group provides more particles that can explore to the left of the resetting position.
Consequently, the probability that at least one particle reaches $x=0$ increases with $N$, reducing $\langle\tau_f\rangle$.
At sufficiently large $N$, this effect dominates the tendency of resetting to shift the group away from the boundary, leading to the observed decrease of $\langle\tau_f\rangle$ with $N$.
An asymptotic analysis predicts that this crossover occurs approximately when
\begin{equation}
    \frac{r/k}{2\ln N^*} \simeq 1,
\end{equation}
for large $N$ (see Appendix~\ref{appendix:c}).


\subsection{Resetting broadens the hitting-time distribution and delays group arrival}\label{sec:3.2}

\begin{figure*}[t!]
 \centering
 \includegraphics[width=\columnwidth]{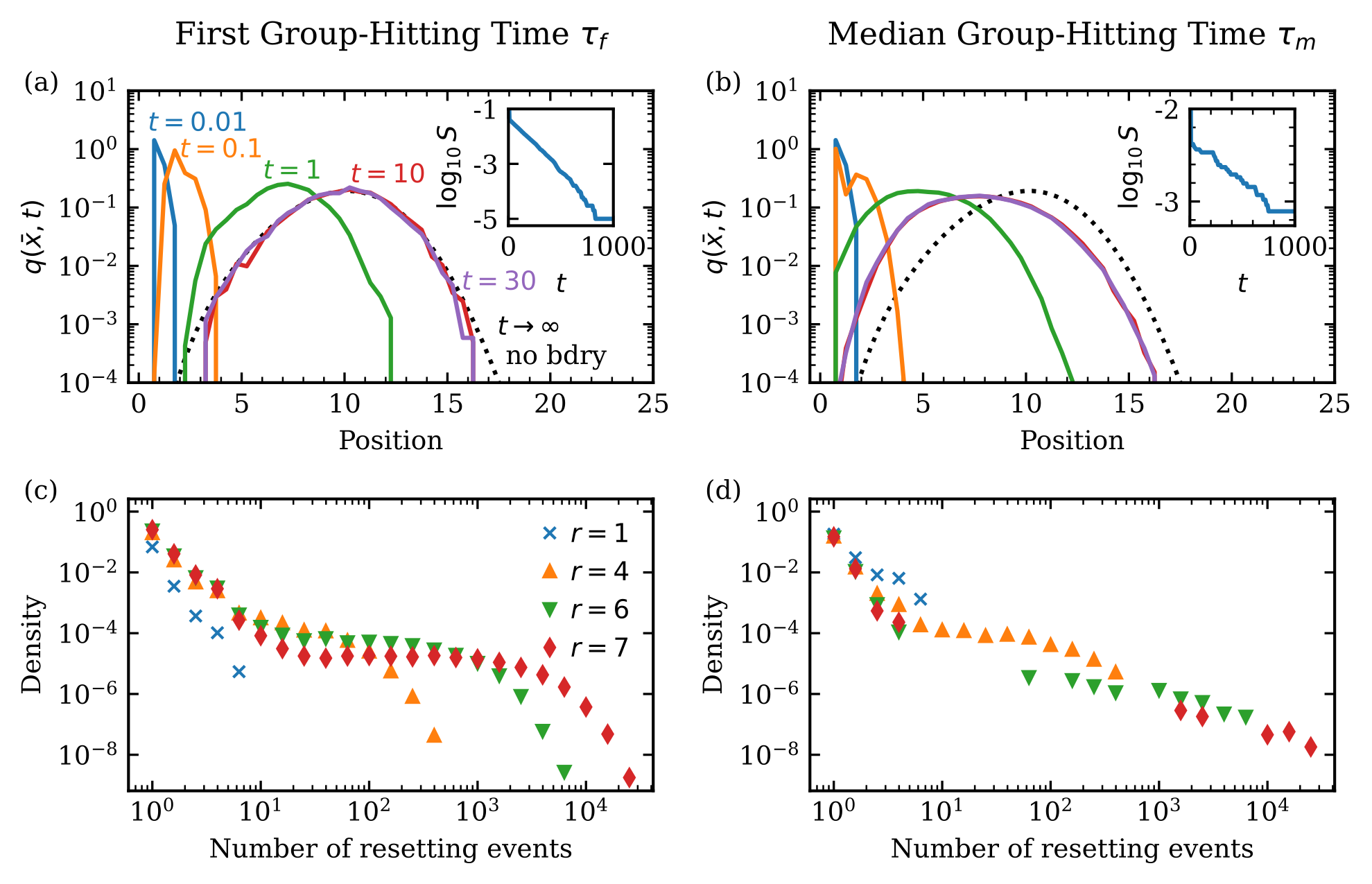}%
 \caption{\label{fig:3}
 Panels (a) and (b): Position distributions of the group mean positions $\bar{x}(t) = \sum_{i} x_i(t) /N$ ($N=100$) for measuring (a) \fght and (b) \mght at different time points, with a resetting rate of $r=6$. The distributions represent the normalized count of surviving cases at each time point. In both cases, the mean position shifts to the right over time due to group resetting, eventually reaching a quasi-stationary state. The black dashed line indicates the stationary state of the case without an adsorbing boundary, which is given in~\cite{lee2026general}. Insets: Log of survival probabilities over time ($t\leq1000$).
 Panels (c) and (d): distributions of the number of resetting events.
 The number of resetting events before adsorption grows with the resetting rate $r$ [(c)\fght  and (d) \mght].
 We note that the number distribution develops into a plateau-like tail.
 }
\end{figure*}
Resetting strongly broadens both the \fght and \mght distributions.
At short times, the distributions closely follow the no-resetting case, with initial peaks occurring at approximately the same times.
At longer times, however, resetting generates broad, plateau-like tails that extend over progressively longer time scales as $r$ increases [Figs.~\ref{fig:2}(a) and (d)].
These long tails arise from realizations that undergo repeated resetting before satisfying the group-hitting criterion and lead to the sharp increase in both $\langle\tau_f\rangle$ and $\langle\tau_m\rangle$ with $r$.

The plateaus suggest that, at long times, the surviving realizations approach a quasi-stationary state in which the probability of satisfying the group-hitting criterion between successive resets becomes small and approximately constant.
To test this hypothesis, we examine the distribution of the group mean position,
$\bar{x}(t)=\sum_i x_i(t)/N$, at different times [Figs.~\ref{fig:3}(a) and (b)].
The distribution over all realizations, $P(\bar{x},t)$, can be written in terms of the survival probability $S(t)$ and the distribution conditioned on survival, $q(\bar{x},t)$, as
\begin{equation}
    P(\bar{x},t)=S(t)q(\bar{x},t),
\end{equation}
where
\begin{equation}
    q(\bar{x},t)=\frac{P(\bar{x},t)}{S(t)},
    \qquad
    \int q(\bar{x},t)\,d\bar{x}=1.
\end{equation}
At short times, $q(\bar{x},t)$ evolves from the initial peak near $x_0$ toward a broader distribution shifted away from the adsorbing boundary [$t=1$ and $t=10$ in Figs.~\ref{fig:3}(a) and (b)].
At longer times, the conditional distribution approaches an approximately time-independent shape,
$q(\bar{x},t)\to q_{\mathrm{QS}}(\bar{x})$ for $t\gg1$.
This quasi-stationary distribution resembles the stationary distribution obtained without an adsorbing boundary~\cite{lee2026general} (black dotted lines), but has depleted flanks because realizations satisfying the group-hitting criterion are removed from the surviving ensemble.
The insets show that this quasi-stationary regime is accompanied by an approximately linear decay of $\log S(t)$, implying
\begin{equation}
    S(t)\sim e^{-\lambda t},\label{eq:Stdecay}
\end{equation}
with an approximately constant escape rate $\lambda$.
Thus, the convergence of $q(\bar{x},t)$ toward $q_{\mathrm{QS}}(\bar{x})$, together with the exponential decay of $S(t)$, shows that the surviving ensemble approaches a slowly decaying quasi-stationary state.

In this quasi-stationary state, there is a small, but finite, probability of satisfying the group-hitting criterion during a given time interval.
This means that surviving realizations can persist for long times and undergo many successive resetting events before the group-hitting criterion is eventually met. 
To test this hypothesis, we record the number of resetting events before the \fght and \mght criteria are met [Figs.~\ref{fig:3}(c) and (d)].
As in the group-hitting-time distributions in Figs.~\ref{fig:2}(a) and (d), increasing $r$ broadens the reset-count distributions and produces plateau-like tails for $r\gtrsim4$.
These plateaus show that the number of resetting events is broad.
For example, at $r=6$, the \fght distribution extends over approximately two orders of magnitude ($10^1$--$10^3$), whereas the \mght distribution is even broader ($10^1$--$10^4$).
For the \mght at $r\gtrsim6$, we also observe a sparsely populated region at intermediate numbers of resetting events ($\sim10^1$--$10^3$), consistent with the corresponding feature in the group-hitting-time distribution in Fig.~\ref{fig:2}(d).
Additional simulations with increased statistics show that this sparsely populated region persists, and not an artifact of limited sampling (Appendix~\ref{appendix:d}).

\subsection{Effects of initial position and potential strength}

\begin{figure*}[t]
 \centering
 \includegraphics[width=\columnwidth]{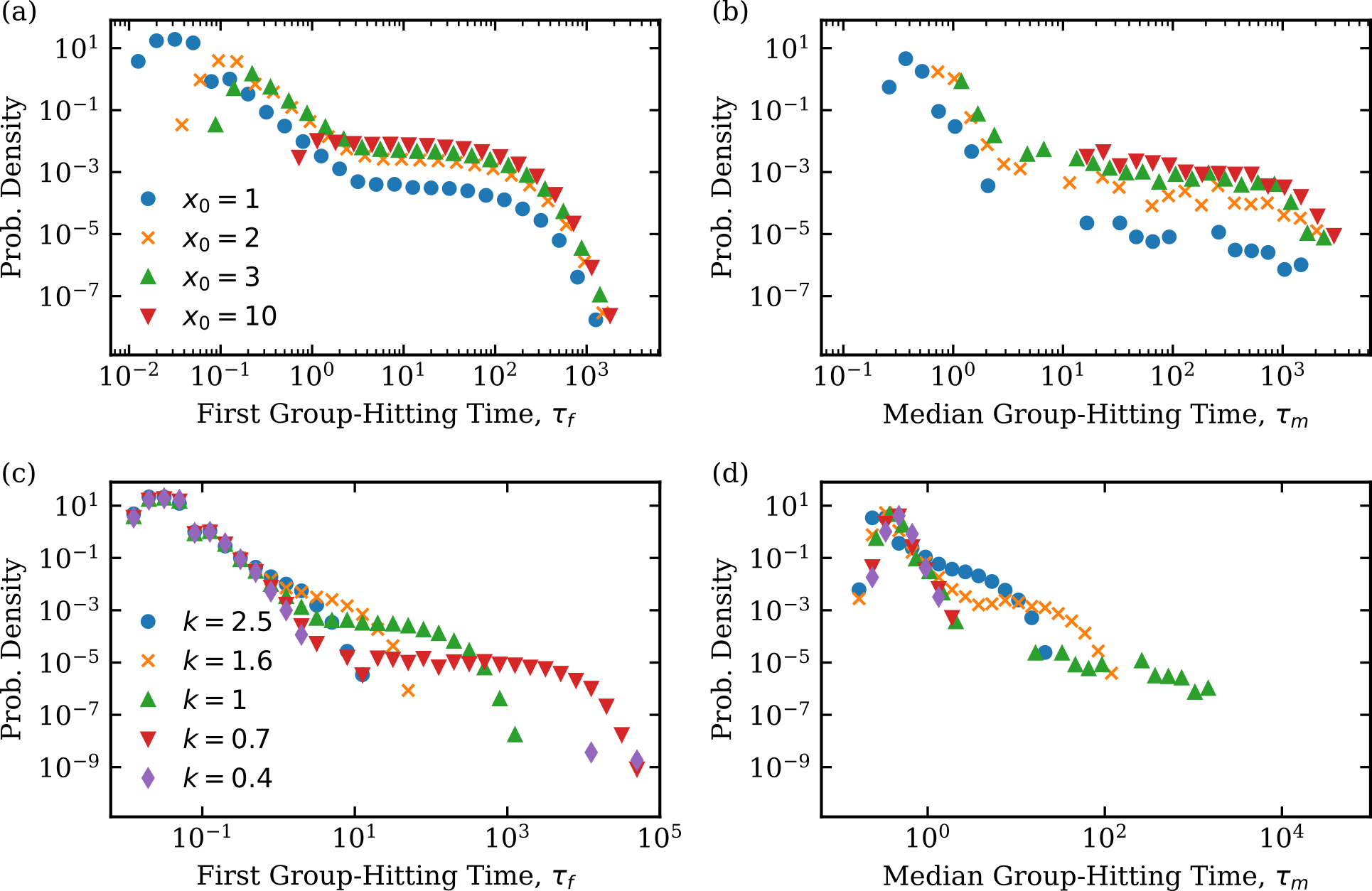}%
 \caption{\label{fig:4} First ($\tau_f$) and median ($\tau_m)$ group-hitting time distributions for $N=100$ and $r=6$ for different initial position $x_0$ [(a) and (b)] and different potential strength $k$ [(c) and (d)]. The other parameters are $D=2,\ r=6$.
}
\end{figure*}

The initial position $x_0$ and potential strength $k$ control different parts of the group-hitting-time distributions: $x_0$ primarily affects the short-time dynamics and the relative weight of the plateau, whereas $k$ controls its long-time extent.
We first examine the dependence on $x_0$ at fixed $r=6$ [Figs.~\ref{fig:4}(a) and (b)].
For the fGHT, increasing $x_0$ shifts the initial peak to later times, lowers its amplitude, and broadens it [Fig.~\ref{fig:4}(a)].
This behavior arises because particles starting farther from the adsorbing boundary generally require more time to reach it, increasing the probability that resetting occurs before the group-hitting criterion is satisfied.
As $x_0$ increases, less probability is therefore concentrated in the short-time peak and more weight appears in the plateau.
After several resets, however, the system loses memory of $x_0$ and approaches the quasi-stationary state.
Consequently, the shape and long-time cutoff of the plateau remain comparatively insensitive to $x_0$.
The \mght distributions exhibit the same qualitative behavior: the early-time peak shifts to later times as $x_0$ increases, whereas the long-time cutoff remains approximately unchanged [Fig.~\ref{fig:4}(b)].

In contrast, the potential strength $k$ primarily controls the long-time behavior of the group-hitting-time distributions.
Changing $k$ has relatively little effect on the initial peak but substantially changes the extent of the plateau [Figs.~\ref{fig:4}(c) and (d)].
Increasing $k$ strengthens the confinement and localizes the quasi-stationary state closer to the adsorbing boundary, thereby increasing the group-hitting rate and shortening the long-time tail.
Conversely, decreasing $k$ allows the quasi-stationary state to extend farther from the boundary, reducing the group-hitting rate and shifting the cutoff to later times.
Thus, $x_0$ primarily controls the short-time dynamics and the relative weight of the plateau, whereas $k$ controls the long-time scale over which the plateau extends.

\subsection{Analytical approximations for hitting-time distributions}

This section develops a mathematical approximation to explain the \fght and \mght distributions. The approximation rests on a renewal structure: at each
reset, all particles relocate to a single position, the rightmost
particle's position $X$, so the subsequent dynamics depends only on $X$.
This allows us to write a renewal equation for the group-hitting-time
density $H^{f,m}(x_0,t)$ (Appendix~\ref{appendix:b}), which we approximate
separately at short and long times. The short-time part is an expansion
in the number of resetting events, and the long-time part is an
exponential decay. We combine the two in Eq.~\eqref{eq:appr} and show the results along with the simulated data in Fig.~\ref{fig:2}(a) and (d) as dashed lines.

Since resets form a Poisson process with rate $r$, the probability of
exactly $n$ resets before time $t$ is $(rt)^n e^{-rt}/n!$.
Hence, at short times, the density is dominated by realizations with few resets.
In particular, the shape of the Group-Hitting Time distribution is largely consistent with the case without resets ($r=0$). 
This suggests that the distribution may be expressed as a series.
\begin{equation}
    H_\text{short}^{f,m}(x_0,t) = \sum_{n=0}^{n_\text{short}}H_n^{f,m}(x_0,t),\label{eq:short}
\end{equation}
where $H_n^{f,m}(x_0,t)$ denotes the group-hitting time density after $n$ resets with starting position $x=x_0$.
The zeroth term is the no-reset
density weighted by the probability that no reset has occurred,
\begin{equation}
  H^{f,m}_0(x_0,t) = e^{-rt}\,\tilde H^{f,m}(x_0,t),
  \label{eq:noreset}
\end{equation}
where $\tilde H^{f,m}$ is given in Eqs.~\eqref{eq:Hf0} and~\eqref{eq:Hm0}. The first-order term reads
\begin{equation}
  H^{f,m}_1(x_0,t) = \int_0^t dT\, re^{-rT}\,\tilde S^{f,m}(x_0,T)
  \int_0^\infty dX\, K_N(X|x_0;T)\,e^{-r(t-T)}\,\tilde H^{f,m}(X,t-T).
  \label{eq:onereset}
\end{equation}
It describes a group that survives until time $T$ [probability $\tilde S^{f,m}(x_0,T)=1-\int_0^TdT'\,\tilde H^{f,m}(x_0,T')$], resets
at $T$ to the rightmost position $X$ [kernel $K_N(X | x_0; T)$], and then meets the
criterion at $t$ without further resets. Following Ref.~\cite{lee2026general}, we approximate $K_N(X | x_0; T)$ by a
Gumbel distribution for large $N$, $\mathrm{Gumbel}(X; \mu, \beta)$, as predicted by the Fisher--Tippett--Gnedenko theorem~\cite{fisher1928limiting,hansen2020three,cartwright1956statistical}. 
The location and scale parameters $\mu = \mu(x_0, T)$ and $\beta = \beta(x_0, T)$ are process-dependent.
A detailed derivation of the approximation $K_N(X | x_0; T) \approx \mathrm{Gumbel}(X; \mu, \beta)$ is provided in Ref.~\cite{lee2026general}.

At long times, many resets have occurred and the surviving realizations
have reached the quasi-stationary state identified in
Sec.~\ref{sec:3.2}. At quasi-stationary, the group meets the hitting criterion at an
approximately constant rate. So the density decays exponentially,
\begin{equation}
  H^{f,m}_{\mathrm{long}}(x_0,t) \sim \frac{e^{-t/\gamma}}{\gamma}.
  \label{eq:long}
\end{equation}
The parameter $\gamma$ is the characteristic time of exponential cut-off.
The mathematical detail is given in  Appendix~\ref{appendix:b}. Rather than computing $\gamma$ explicitly, we determine $\gamma$ from the simulated mean $\langle\tau_{f,m}\rangle$.

Combining the short- and long-time contributions, we obtain the following
approximate formula for the \fght and \mght 
\begin{align}
\begin{split}
&H^{f,m}(x_0,t) = H_\text{short}^{f,m}(x_0,t) + H_\text{long}^{f,m}(x_0,t)\\ 
& \approx H_0^{f,m}(x_0,t)+H_1^{f,m}(x_0,t)+(1-q_0-q_1)\frac{e^{-t/\gamma}}{\gamma},
\label{eq:appr}
\end{split}
\end{align}
where 
\begin{equation}
    q_0=\int_0^\infty H_0^{f,m}(x_0,t)dt \ \ \ \mathrm{and} \ \ \ q_1=\int_0^\infty H_1^{f,m}(x_0,t)dt
\end{equation}
For simplicity, we kept contributions up to one resetting event in Eq.~\eqref{eq:appr} in addition to the exponential cutoff.

In Figs.~\ref{fig:2}(a) and (d), we show the approximate \fght and \mght distributions.
We evaluate Eq.~\eqref{eq:onereset} numerically and use the values of $\langle\tau_{f,m}\rangle$ obtained from the simulations ($\gamma=\langle\tau_{f,m}\rangle$).
The approximations show qualitative agreement with the simulations for the \fght, whereas the agreement is poorer for the \mght because $\langle\tau_{f,m}\rangle$ is a lower bound obtained from right-censored data.
In particular, the approximation captures the emergence of the plateau at large $r$, which reflects strong heterogeneity among realizations.

\subsection{Group-hitting times become increasingly heterogeneous}

\begin{figure*}[t!]
    \centering
    \includegraphics[width=\columnwidth]{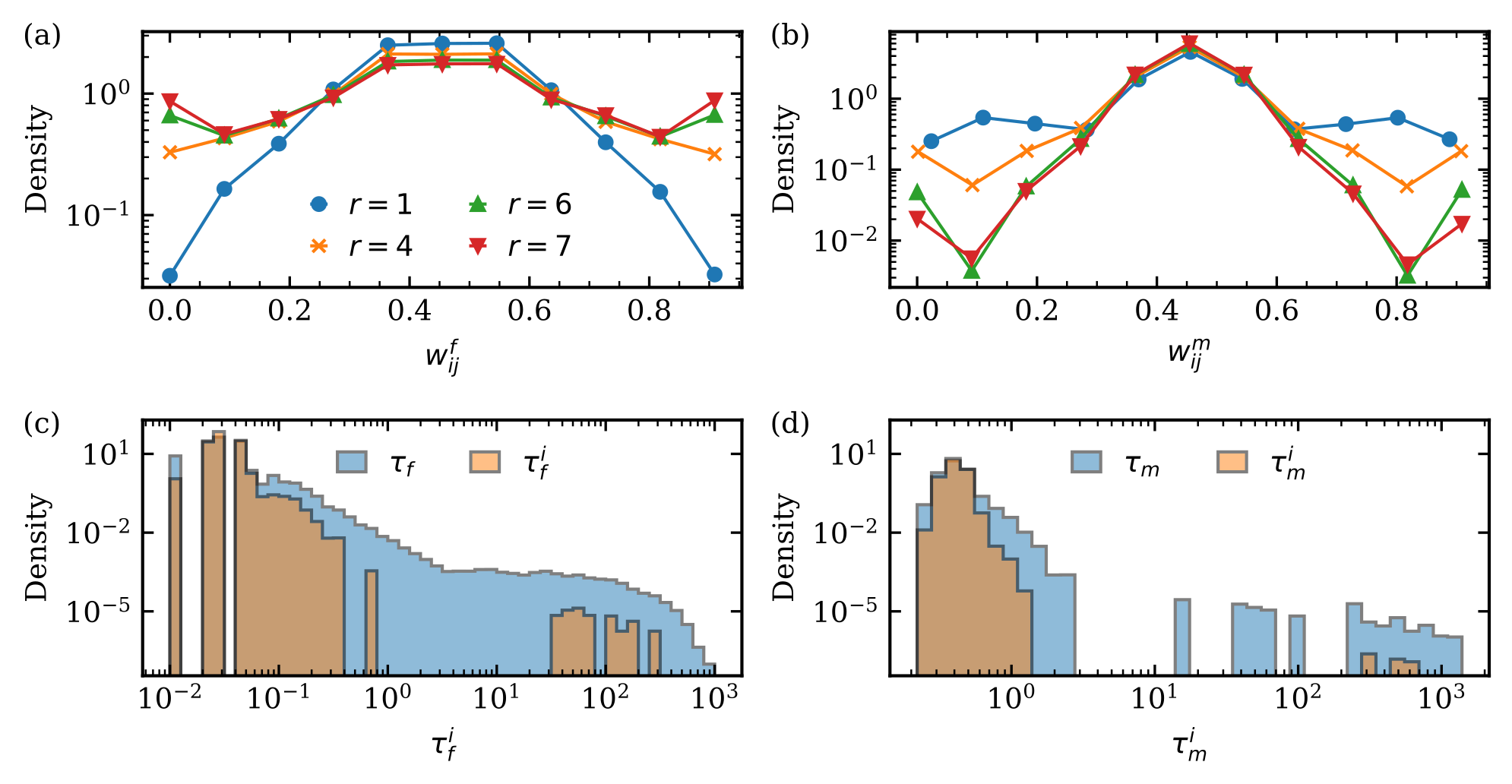}%
    \caption{\label{fig:5}
    The distribution of the uniformity index $w^{f,m}_{ij}$ for \fght (a) and \mght (b).
    The flanks become more pronounced with increasing $r$, indicating that realizations become more dissimilar ($N=100$).
    (c,d) The distribution of $\tau^i_{f,m}$ and $\tau^j_{f,m}$ yielding $0.45<w^{f,m}_{ij}<0.55$ for $r=6$ (orange).   
    Blue shaded areas show the histogram of all group hitting times $\tau_{f,m}$.
    }
\end{figure*}

In Fig.~\ref{fig:2}, we observe that the \fght and \mght distributions broaden significantly as $r$ increases.
This suggests that the averages $\langle \tau_f\rangle$ and $\langle \tau_m\rangle$ do not represent typical group-hitting times.
To examine this hypothesis, we introduce the uniformity index $w_{ij}$ \cite{grebenkov2018strong, mattos2012first, hedstrom2024enhancer}, which quantifies the relative hitting times of two different realizations, $i$ and $j$, defined as
\begin{equation}
    w_{ij}^{f,m} = \frac{\tau_{f,m}^i}{\tau_{f,m}^i + \tau_{f,m}^j}.
\end{equation}
Here, $\tau_{f,m}^i$ denotes the group-hitting time of realization $i$.
Similar group-hitting time pairs give $w_{ij}^{f,m}\approx0.5$, whereas strongly dissimilar times give values close to $0$ or $1$.
Thus, a bell-shaped distribution centered around $0.5$ indicates relatively homogeneous group-hitting times, whereas increasing weight near the flanks indicates greater heterogeneity.

We show the uniformity-index distributions for different resetting rates in Figs.~\ref{fig:5}(a) and (b).
For each $r$, we randomly select $10^5$ realization pairs (excluding those with group-hitting times exceeding the simulation time $\tau_{f,m}^i\geq t_{\mathrm{max}}$).
At $r=1$ (blue), both the \fght and \mght distributions are bell-shaped and peaked at $w_{ij}^{f}=w_{ij}^{m}=0.5$.
The left and right flanks are weak, indicating that strongly dissimilar pairs are rare.
Thus, at low $r$, most realizations have similar group-hitting times.

As $r$ increases, the flanks ($w_{ij}^{f,m}\approx0$ and $1$) become more pronounced relative to the shoulders ($w_{ij}^{f,m}\approx0.25$ and $0.75$), indicating greater heterogeneity among realizations.
For $r\gtrsim4$, the increasing flank weight marks the onset of strong heterogeneity, which becomes more pronounced as $r$ increases.
In this regime, the mean group-hitting time becomes less representative of individual realizations.
For the mGHT, the absolute height of the flanks decreases at large $r$ because we exclude realizations with $\tau_m>t_{\mathrm{max}}$, thus removing the longest group-hitting times [Fig.~\ref{fig:5}(b)].

To identify the group-hitting times that contribute to the central peak, we examine pairs satisfying $0.45<w_{ij}^{f,m}<0.55$ at $r=6$ [orange histograms in Figs.~\ref{fig:5}(c) and (d)].
These pairs contain both short and long group-hitting times, showing that $w_{ij}^{f,m}\approx0.5$ does not necessarily correspond to a single characteristic time scale.
Pairs of similarly short or similarly long times contribute to the central peak, whereas pairs of strongly dissimilar times contribute to the flanks.
The complete group-hitting-time distributions are shown in blue for comparison.

\section{Discussion and Conclusion}
This paper studies the first-passage properties of a many-particle system in which the resetting protocol depends on the internal dynamics of the group.
Building on our previous work motivated by artificial group selection, we consider particles diffusing in a confining potential toward an adsorbing boundary, with the entire group reset to the position of the rightmost particle.

Unlike single-particle systems, many-particle systems open for several definitions of first passage.
Here, we consider two: the first group-hitting time (\fght), when any particle reaches the boundary, and the median group-hitting time (\mght), when half of the particles have reached it.
Other group-hitting criteria, such as the time at which all particles reach the boundary, or a randomly selected particle reaches it, could lead to different collective first-passage behavior and remain interesting directions for future work.

The shape of our group-hitting-time distributions---a pronounced short-time peak followed by a broad plateau and a long-time cutoff---resembles the reaction-time distribution studied in Ref.~\cite{grebenkov2018strong}.
That study considered a diffusing reactant searching for an imperfect reaction center represented by a partially reflecting sphere.
The peak of the distribution is associated with direct search paths and immediate reactions, whereas the plateau emerges when the reactant diffuses around the reaction center and repeatedly attempts to react. The reaction-time distribution exhibits two relevant time scales: the most probable reaction time (the peak) and the mean reaction time. As the reaction rate constant decreases, these two separate, making it difficult to assign a typical reaction time. This regime is referred to as heterogeneity-controlled kinetics \cite{godec2016first}. Mapping these parameters to our system, we find that the resetting rate plays a role analogous to the inverse imperfect reaction rate. Increasing the resetting rate extends the plateau in our case, whereas lowering the reaction rate extends it in \cite{grebenkov2018strong}.

In conclusion, our simulations show that the first and median group-hitting-time distributions have the same general shape: a short-time peak, close to that of the no-resetting case, followed by a plateau and an exponential long-time decay.
This plateau spans several orders of magnitude, especially at high resetting rates.
We demonstrated that the extended plateau and subsequent exponential decay arise from a quasi-stationary state with an approximately constant escape rate associated with satisfying the group-hitting criterion between resetting events.
The mean group-hitting times also show similar dependence on the resetting rate, increasing sharply at large $r$.
Their dependence on group size, however, is opposite: for sufficiently large $N$, the mean \fght decreases with $N$, whereas the mean \mght increases.
In both cases, the first-passage dynamics become increasingly heterogeneous as the resetting rate increases.
Consequently, the mean becomes less representative of individual realizations, demonstrating the importance of looking beyond average first-passage times to characterize heterogeneous search and avoidance dynamics in many-particle systems with resetting.

%
%
\begin{acknowledgments}
This work is supported by the Swedish Research Council (Grants No.~2021-04080 and No.~2022-06543) and the Swedish Foundation for International Cooperation in Research and Higher Education (STINT) (Grant No.~MG2022-9405).
S.-G.Y. is supported by the Basic Science Research Program through the National Research Foundation of Korea (NRF) funded by the Ministry of Education (Grant No.~RS-2025-02312897).
J.L. acknowledges support from the Kempestiftelserna (Grant No.~JCK22-0026.3).
\end{acknowledgments}

\appendix
\section{Group-hitting-time distributions without resetting ($r=0$)}\label{appendix:a}
\setcounter{equation}{0}
\setcounter{secnumdepth}{2}
\renewcommand{\theequation}{A\arabic{equation}}
In this appendix section, we derive the first- and median- group hitting time density with no resets, $\hat {\tilde{H}}^{f,m}(x_0,t)$. Without resets, the particles move independently, so the group-hitting times are the order statistics of $N$ independent single-particle first-passage times~\cite{weiss1983order}: the \fght is the smallest of them, and the \mght is the $M$-th smallest, with
$M=\lfloor N/2\rfloor$. To obtain the expression, we need the first-hitting-time density of a single particle in the harmonic potential $V(x)=\frac12kx^2$, denoted $\tilde{H}^{s}(x_0,t)$. 
When the boundary is at $x=a<x_0$, its Laplace-transformed solution $\hat{H}^{s}(x_0,s) = \int_0^\infty e^{-st}\tilde{H}^s(x_0,t) dt$ is~\cite{siegert1951first,alili2005representations}
\begin{equation}
\hat{\tilde H}^{s}(x_0,s) = \frac{e^{kx_0^2/4D}D_{-s/k}(x_0\sqrt{k/D})}{e^{ka^2/4D}D_{-s/k}(a\sqrt{k/D})},
\label{eq:lap}
\end{equation}
where $D_{-\nu}(\cdot)$ is a parabolic cylinder function with $\nu=s/k$ in our case.
The inverse Laplace transform of~\eqref{eq:lap} is known for $a=0$ and reads~\cite{alili2005representations}
\begin{equation}
\tilde{H}^{s}(x_0,t) = \frac{x_0}{2\sqrt{\pi D}}\left(\frac{k}{\sinh kt}\right)^{\frac32}\exp\left(-\frac{kx_0^2e^{-kt}}{4D\sinh kt}+\frac{kt}{2}\right).
\label{eq:inv}
\end{equation}

Following Ref.~\cite{weiss1983order}, the \fght density $\tilde{H}^{f}(x_0,t)$ follows from requiring that one particle hits the
boundary at $t$ while the remaining $N-1$ particles survive, defined as
\begin{equation}
\tilde{H}^f(x_0,t)=N \tilde H^{s}(x_0,t) \tilde{S}^s(x_0,t)^{N-1},\label{eq:Hf0}
\end{equation}
where $\tilde{H}^{s}(x_0,t)$ is the first-hitting time density for when a single particle starting at $x=x_0$ hits the boundary at $t$ without reset.
$\tilde{S}^s(x_0,t)=1-\int_0^t \tilde{H}^s(x_0,T)dT$ is the survival probability at time $t$.

Similarly, the \mght density follows from requiring that one particle hits the boundary at $t$, $k-1$ particles have already been absorbed,
and the remaining $N-k$ particles survive,  
\begin{equation}
\tilde{H}^m(x_0,t)=\left\lfloor \frac{N}{2}\right\rfloor \begin{pmatrix}N\\ \lfloor \frac{N}{2}\rfloor\end{pmatrix}\tilde{H}^s(x_0,t)[\tilde{S}^s(x_0,t)]^{N-\left\lfloor \frac{N}{2}\right\rfloor}\left[1-\tilde{S}^s(x_0,t)\right]^{\left\lfloor \frac{N}{2}\right\rfloor-1}.
\label{eq:Hm0}
\end{equation}
%


\section{Derivation of the approximate group-hitting-time distribution}\label{appendix:b}
\setcounter{equation}{0}
\setcounter{secnumdepth}{2}
\renewcommand{\theequation}{B\arabic{equation}}
In this appendix, we derive the two ingredients of Eq.~\eqref{eq:appr}.
We first write a renewal equation for $H^{f,m}(x_0,t)$ (Section \ref{app:renewal}).
Expanding it in the number of resetting events yields the short-time
terms $H^{f,m}_0$ and $H^{f,m}_1$ (Section \ref{app:short}).
Analyzing its Laplace transform shows that the long-time decay is
exponential with a strictly positive rate $1/\gamma = r - s_0$ (Section \ref{app:long})

\subsection{Renewal equation and approximation}\label{app:renewal}
Each reset relocates all particles to a single position $X$, so the
dynamics after a reset depends only on $X$. Conditioning on the first
resetting event, following Ref.\cite{dahlenburg2021stochastic}, gives
\begin{equation}
  H^{f,m}(x_0,t) = \Psi(t)\,\tilde H^{f,m}(x_0,t)
  + \int_0^t dT\,\phi(T) \int_0^\infty dX\,
    \tilde S^{f,m}(x_0,T)\, K_N(X|x_0;T)\, H^{f,m}(X,t-T),
  \label{eq:renewal}
\end{equation}
with $\Psi(t)=e^{-rt}$ and $\phi(t)=re^{-rt}$.
The first term represents a case that no reset occurs before $t$ from the beginning.
The second term describes a situation at the first reset event at $T$, the group has not met the hitting criterion by then, the particles relocate to $X$, and the process restarts from $X$. This term is composed by a product of the waiting time between resetting $\phi(t)$, the survival probability of the group (with no reset) $\tilde S^{f,m}(x_0,t)$, and the kernel describing the maximum-position resetting process $K_N(x|x_0;t)$~\cite{lee2026general}.
To keep the analysis tractable, we assume that $K_N(X|x_0;T)$ is negligibly small for $X<0$, so that the integration over $X$ can be restricted to $X\geq0$.

\subsection{Short-time behavior}\label{app:short}
Iterating Eq.~\eqref{eq:renewal} gives $H^{f,m} = \sum_{n=0}^\infty H^{f,m}_n$, where $H^{f,m}_n$ is the
contribution from exactly $n$ resetting events, 
\begin{align}
  H^{f,m}_0(x_0,t) &= \Psi(t)\,\tilde H^{f,m}(x_0,t), \\
  H^{f,m}_n(x_0,t) &= \int_0^t dT\,\phi(T)\int_0^\infty dX\,
    \tilde S^{f,m}(x_0,T)\,K_N(X|x_0;T)\,H^{f,m}_{n-1}(X,t-T),
    \quad n\ge 1 .
  \label{eq:B-recursion}
\end{align}
For $n=0$ and $n=1$, these reduce to Eqs.~\eqref{eq:noreset} and ~\eqref{eq:onereset}.
Since resets form a Poisson process, the probability of exactly $n$ resets before time $t$ is $(rt)^n e^{-rt}/n!$, so terms with small $n$ dominate for $t \lesssim 1/r$. Consistently, the reset-count distributions peak at small $n$ [Figs.~\ref{fig:3}(c) and (d)]. Terms with larger $n$ contribute at later times. In Eq.~\eqref{eq:appr}, we keep
$n\le 1$ explicitly and represent the remainder by the exponential tail
derived next.

\subsection{Long-time behavior}\label{app:long}
Writing $H^{f,m}(x_0,t) = e^{-rt}\,\check H^{f,m}(x_0,t)$   the
reset-waiting factors, so that
\begin{equation}
  \check H^{f,m}(x_0,t) = \tilde H^{f,m}(x_0,t)
  + r\int_0^t dT\int_0^\infty dX\,
    \tilde S^{f,m}(x_0,T)\,K_N(X|x_0;T)\,\check H^{f,m}(X,t-T). \label{eq:renewal2}
\end{equation}
The Laplace transform of Eq.~\eqref{eq:renewal2} is
\begin{equation}
\hat{\check H}^{f,m}(x_0,s)=\hat{\tilde H}^{f,m}(x_0,s)+r\int_0^\infty dX
\ W^{f,m}(X|x_0;s)\hat{\check H}^{f,m}(X,s)\label{eq:LT_renew2}
\end{equation}
where $W^{f,m}(X|x;s) = \int_0^\infty dt\, e^{-st}\,\tilde S^{f,m}(x,t)\,K_N(X|x;t)$.
Define the integral operator
\begin{equation}
  (\mathcal W_s f)(x) = \int_0^\infty dX\, W^{f,m}(X|x;s)\, f(X).
\end{equation}
Then the solution is 
$\hat{\check H}^{f,m} = [\mathcal I - r\mathcal W_s]^{-1}\hat{\tilde H}^{f,m}$. The long-time behavior is set by the rightmost singularity $s_0$ of the
resolvent, where the Perron eigenvalue $\Lambda(s)$ of $r\mathcal W_s$
satisfies $\Lambda(s_0)=1$. Hence $\check H \sim e^{s_0 t}$ and
\begin{equation}
  H^{f,m}(x_0,t) \sim e^{-t/\gamma}, \qquad \gamma = \frac{1}{r-s_0}.
\end{equation}

For $H^{f,m}$ to decay, the pole must satisfy $s_0<r$. We now show that this is always the case. Let $\theta>0$ denote the decay rate of the no-reset survival probability, $\tilde S^{f,m}(x,t)\sim e^{-\theta t}$. Assume that $r\mathcal W_s$ is positive for
$s>-\theta$, so that its Perron eigenvalue $\Lambda(s)$ is well defined.
We then show that $\Lambda(s)$ crosses unity somewhere in $(-\theta,r)$.
 
We first evaluate $\Lambda$ at $s=r$. Acting with $r\mathcal W_s$ on the
constant function $\mathbf 1$ eliminates the kernel $K_N$, since
$\int_0^\infty K_N\,dX \le 1$, leaving
$(r\mathcal W_s\mathbf 1)(x) \le r\hat{\tilde S}^{f,m}(x,s)$.
At $s=r$, the right-hand side equals $1-\hat{\tilde H}^{f,m}(x,r)$,
which is strictly smaller than one for every $x$. Pairing this
inequality with the positive left Perron eigenfunction $\psi_r$ gives
$\Lambda(r)\langle\psi_r,\mathbf 1\rangle
  = \langle\psi_r, r\mathcal W_r\mathbf 1\rangle
  < \langle\psi_r,\mathbf 1\rangle$,
and hence $\Lambda(r)<1$.
 
Next, because the kernel of $\mathcal W_s$ carries the factor $e^{-sT}$,
$\Lambda(s)$ decreases monotonically with $s$. In the opposite limit
$s\rightarrow-\theta$, the Laplace transform $\hat{\tilde S}^{f,m}(x,s)$
diverges, and so does $\Lambda(s)$.

Since $\Lambda(s)$ is continuous, decreasing, unbounded as
$s\rightarrow-\theta$, and smaller than one at $s=r$, the equation
$\Lambda(s_0)=1$ has a solution $s_0\in(-\theta,r)$. Therefore
$1/\gamma = r-s_0>0$, and the long-time tail always decays
exponentially.

The rate $1/\gamma = r-s_0$ is the escape rate $\lambda$ of the
quasi-stationary state in Eq.~\eqref{eq:Stdecay}, consistent with the linear decay of
$\log S(t)$ in the insets of Fig.~\ref{fig:3}.
In the main text, we do not compute $s_0$ explicitly; instead, $\gamma$
in Eq.~\eqref{eq:appr} is fitted to $\langle\tau_{f,m}\rangle$

\section{Asymptotic dependence of the first group-hitting time on group size $N$\label{appendix:c}}
\setcounter{equation}{0}
\setcounter{secnumdepth}{2}
\renewcommand{\theequation}{C\arabic{equation}}
In this appendix, we provide an approximate description of how $\langle\tau_f\rangle$ depends on $N$ at large $r$. To this end, we assume that the system has reached a quasi-stationary state, in which the resetting position can be approximated by a fixed stationary resetting position $x_s$. For $x_s$, we borrow the mean stationary resetting position derived in~\cite{lee2026general} for the same resetting process without an adsorbing boundary. The result is intended to identify the scale at which the crossover in $N$ occurs, not to predict it quantitatively. We proceed in three steps. We first derive an equation for the mean first group-hitting time and solve it with a fixed
resetting position. We then evaluate the solution asymptotically for large $r$ and large $N$. Finally, we identify the crossover.

We first derive an equation for the mean first group-hitting time
$\langle\tau_f\rangle(x_0)$. Taking the Laplace
transform of the renewal equation~(B1) for the \fght, with
$\Psi(t)=e^{-rt}$ and $\phi(t)=re^{-rt}$, gives
\begin{equation}
  \hat H^f(x_0,s) = \hat{\tilde H}^f(x_0,s+r)
  + r\int_0^\infty dX\, W^f(X|x_0;s+r)\,\hat H^f(X,s),
  \label{eq:renewalLaplace}
\end{equation}
where
$W^f(X|x_0;s)=\int_0^\infty dT\,e^{-sT}\tilde S^f(x_0,T)K_N(X|x_0;T)$.
The mean follows from $\langle\tau_f\rangle(x_0)=-\partial_s\hat H^f(x_0,s)|_{s=0}$.
Differentiating Eq.~\eqref{eq:renewalLaplace} with respect to $s$, setting
$s=0$, and using $\hat H^f(X,0)=1$, we obtain
\begin{equation}
  \langle\tau_f\rangle(x_0) = -\partial_s\hat{\tilde H}^f(x_0,r)
  - r\int_0^\infty dX\,\partial_s W^f(X|x_0;r)
  + r\int_0^\infty dX\, W^f(X|x_0;r)\,\langle\tau_f\rangle(X),
  \label{eq:renewalLapderiv}
\end{equation}
where $\partial_s$ acts on the last argument. The first two terms
simplify as follows. Because $\int_0^\infty K_N\,dX=1$ (neglecting the weight at $X<0$, as in Appendix~\ref{appendix:b}), integrating $W^f$ over
$X$ gives
$\int_0^\infty dX\,W^f(X|x_0;s)=\hat{\tilde S}^f(x_0,s)$, and hence
$\int_0^\infty dX\,\partial_sW^f=\partial_s\hat{\tilde S}^f$.
Moreover, $\hat{\tilde H}^f(x_0,s)=1-s\hat{\tilde S}^f(x_0,s)$ implies $\partial_s\hat{\tilde H}^f=-\hat{\tilde S}^f-s\,\partial_s\hat{\tilde S}^f$.
Substituting both relations into Eq.~\eqref{eq:renewalLapderiv}, the terms
containing $\partial_s\hat{\tilde S}^f$ cancel, leaving
\begin{equation}
  \langle\tau_f\rangle(x_0) = \hat{\tilde S}^f(x_0,r)
  + r\int_0^\infty dX\, W^f(X|x_0;r)\,\langle\tau_f\rangle(X).
  \label{eq:renewalmean}
\end{equation}
This equation has a direct interpretation. The first term is the mean time until either the group hits the boundary or the first reset occurs. The second term accounts for the case in which the reset comes first, which happens with probability $r\hat{\tilde S}^f(x_0,r)=1-\hat{\tilde H}^f(x_0,r)$, after which the process restarts from the new position $X$.

In the quasi-stationary state, we replace the resetting position by
its mean, $W^f(X|x_0;r)\approx\hat{\tilde S}^f(x_0,r)\,\delta(X-x_s)$,
which preserves the total weight $\int dX\,W^f=\hat{\tilde S}^f$.
Equation~\eqref{eq:renewalmean} then becomes
\begin{equation}
  \langle\tau_f\rangle(x_0) \approx \hat{\tilde S}^f(x_0,r)
  + r\hat{\tilde S}^f(x_0,r)\,\langle\tau_f\rangle(x_s),
  \label{eq:renewalLapderivdelta}
\end{equation}
where $\hat{\tilde S}^f(x_0,r)$ is the probability that the first reset
occurs before the group hits the boundary. Setting $x_0=x_s$ and
solving for $\langle\tau_f\rangle(x_s)$ gives
\begin{equation}
  \langle\tau_f\rangle(x_s)=\frac{\hat{\tilde S}^{f}(x_s,r)}{1-r\hat {\tilde S}^f(x_s,r)}=\frac{1}r\left[\frac{1}{\hat{\tilde H}^f(x_s,r)}-1\right]\approx\frac{1}{r\hat{\tilde H}^f(x_s,r)}
  \label{eq:tauf_sol}
\end{equation}
where the last step assumes that hitting before a reset is rare,
$\hat{\tilde H}^f(x_s,r)\ll1$.

Next, we evaluate $\hat{\tilde H}^f(x_s,r)$. Since
$\tilde S^f=[\tilde S^s]^N$,
\begin{equation}
  \hat{\tilde H}^f(x_s,r) = 1 - r\int_0^\infty dt\,e^{-rt}[\tilde S^s(x_s,t)]^N
  \approx N\,\hat{\tilde H}^s(x_s,r),
  \label{eq:Hflinear}
\end{equation}
where we linearized
$[\tilde S^s]^N\approx 1-N(1-\tilde S^s)$, which is valid when a single
particle rarely hits the boundary before a reset. The single-particle
transform follows from Eq.~\eqref{eq:lap} with $a=0$ and $s=r$. Writing
$\nu=r/k$ and $\tilde x_s = x_s\sqrt{k/D}$, and using the integral
representation
$I(z)=\int_0^\infty dt\,t^{\nu-1}e^{-t^2/2-zt}=\Gamma(\nu)e^{z^2/4}D_{-\nu}(z)$
, we obtain
\begin{equation}
  \hat{\tilde H}^s(x_s,r) = \frac{I(\tilde x_s)}{I(0)} .
\end{equation}
For large $\nu$, we evaluate $I(z)$ by the saddle-point method. The exponent $\psi(t)=\nu\ln t - zt - t^2/2$ is maximal at
$t^*(z)=\bigl(-z+\sqrt{z^2+4\nu}\bigr)/2$, and $I(z)\approx\sqrt{2\pi/(\nu+t^{*2})}\,e^{\psi(t^*)}$.
At large $r$ and $N$, the resetting position scales as $x_s\sim\sqrt{Dr\ln N}/k$ [25], so $\tilde x_s\sim\sqrt{\nu\ln N}$.
When $\tilde x_s^2\gg4\nu$, the saddle point simplifies to
$t^*\approx\nu/\tilde x_s$, while $t^*(0)=\sqrt\nu$. To leading order in $\ln N$, this gives
\begin{equation}
  \hat{\tilde H}^s(x_s,r) \sim (e\ln N)^{-\nu/2},
  \label{eq:HsscaleN}
\end{equation}
up to a prefactor of order one.

Finally, substituting Eqs.~\eqref{eq:Hflinear} and \eqref{eq:HsscaleN}
into Eq.~\eqref{eq:tauf_sol} gives
\begin{equation}
  \langle\tau_f\rangle(x_s) \sim \frac{(e\ln N)^{\nu/2}}{rN},
  \qquad
  \frac{d\ln\langle\tau_f\rangle}{d\ln N} \sim -1 + \frac{r/k}{2\ln N}.
  \label{eq:taufresult}
\end{equation}
Thus, $\langle\tau_f\rangle$ increases with $N$ for $\ln N<r/2k$ and decreases beyond it, which gives the crossover condition in Eq.~(3).
For $r=6$ and $k=1$, this yields $N^*\approx20$.
Since the convergence to the asymptotic scaling is slow at moderate $N$, these values tend to overestimate the crossover observed in simulations. The estimate should thus be understood as explaining the origin of the non-monotonic $N$-dependence, rather than predicting the location of its maximum.

Equation~\eqref{eq:taufresult} describes a system starting from the quasi-stationary resetting position $x_s$. In the simulations, the particles start near the boundary ($x_0=1$). Equation~\eqref{eq:renewalLapderivdelta}
then gives 
$\langle\tau_f\rangle(x_0)\approx\hat{\tilde S}^f(x_0,r)
 + P_{\mathrm{surv}}(x_0)\,\langle\tau_f\rangle(x_s)$.
The fraction $1-P_{\mathrm{surv}}$ of realizations absorbed before the first reset increases with $N$, so $P_{\mathrm{surv}}$ decreases. This additional decrease shifts the crossover to smaller $N$ than predicted by Eq.~\eqref{eq:taufresult}.

\section{Supplementary simulations of the sparse region in the median group-hitting-time distribution at large $r$\label{appendix:d}}
\setcounter{equation}{0}
\setcounter{secnumdepth}{2}
\renewcommand{\theequation}{D\arabic{equation}}
\setcounter{figure}{0}
\renewcommand{\thefigure}{D\arabic{figure}}

\begin{figure}[!h]
\includegraphics[width=0.9\linewidth]{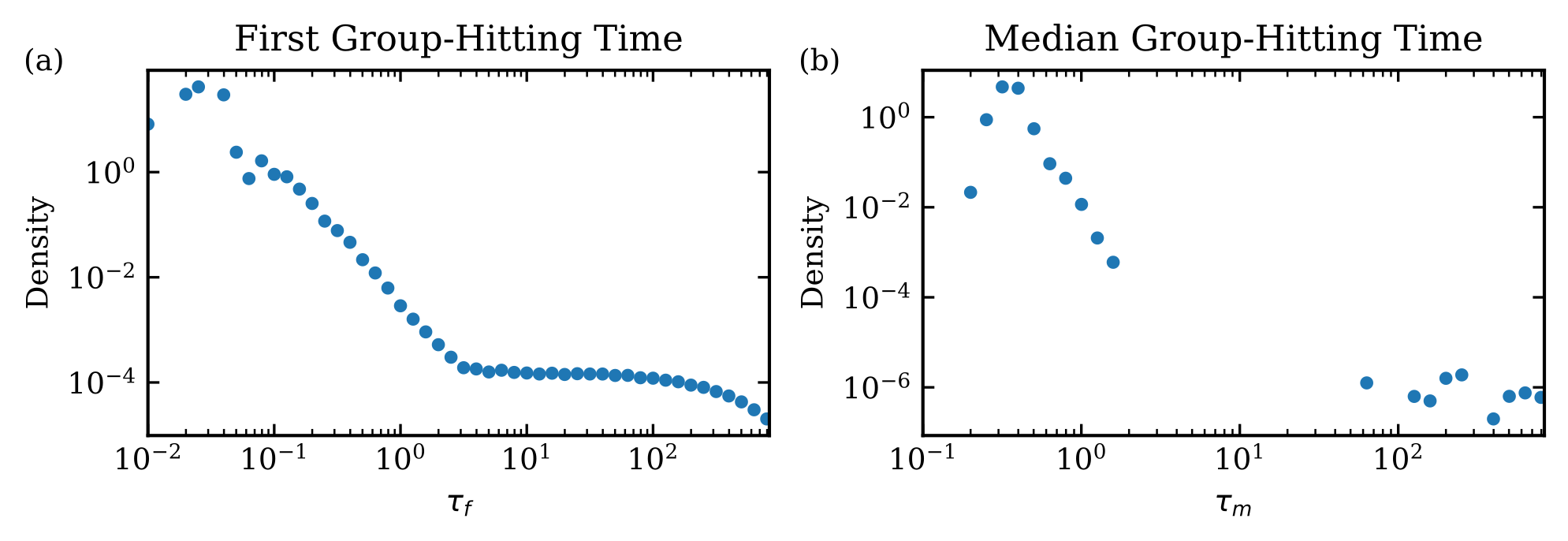}

\caption{Group-hitting-time distributions from $10^7$ realizations for $r=6$ and $N=100$: (a) first group-hitting time and (b) median group-hitting time. Trajectories are terminated once the hitting time exceeds $10^3$.\label{fig:fillgap}}
\end{figure}

This appendix presents supplementary simulation results. We performed simulations with a hundredfold larger sample size, terminating trajectories once $\tau_m>10^3$.(Fig.~\ref{fig:fillgap}). Notably, this figure shows that only about 15,000 out of $10^7$ realizations yield $\tau_m<10^3$.

For the mGHT at $r\gtrsim6$, the distribution in Fig.~\ref{fig:2}(d) shows a
sparsely populated region between the early-time peak and the plateau,
around $\tau_m\sim10^1$. A corresponding region appears in the reset-count distribution in Fig.~\ref{fig:3}(d). Here, we test whether this feature is an artifact of limited sampling.

We repeated the simulations for $r=6$ and $N=100$, with the other
parameters as in Sec.~2.1, using a hundredfold larger sample of $10^7$
realizations. To reduce the computational cost, we terminated each trajectory once $\tau_{f,m}>10^3$.

Figure~\ref{fig:fillgap} shows the resulting distributions. For the fGHT
[Fig.~\ref{fig:fillgap}(a)], the early-time peak connects smoothly to the
plateau. For the mGHT [Fig.~\ref{fig:fillgap}(b)], the sparse region persists: only about 15{,}000 of the $10^7$ realizations reach $\tau_m<10^3$, and none of them fall between the early-time peak and the plateau.
Increasing the sample size by two orders of magnitude therefore does not fill this region, indicating that the gap reflects an intrinsically low probability of intermediate median group-hitting times rather than insufficient sampling.

%
\bibliographystyle{unsrt}
\bibliography{ref.bib}
\end{document}